\documentclass[screen,manuscript,nonacm,review=false,timestamp=false]{acmart}
\pdfoutput=1

\usepackage[normalem]{ulem}
\usepackage[T1]{fontenc}
\usepackage[utf8]{inputenc}
\usepackage[english]{babel}
\usepackage{todonotes}
\usepackage{textcomp}

\usepackage[detect-all]{siunitx}
\usepackage{comment}
\usepackage{algorithm}
\usepackage{algpseudocode}
\usepackage{balance}
\usepackage{mathtools}
\usepackage{float}
\usepackage{blindtext}
\usepackage{caption}
\usepackage{subcaption}
\usepackage{pifont}

\usepackage{setspace} 
\usepackage{wrapfig}
\usepackage{mdframed}
\usepackage{soul}
\usepackage{booktabs}
\usepackage{wrapfig}
\usepackage{balance}

\usepackage[draft=true]{minted}

\usepackage{tikz}

\usepackage{calrsfs}
\DeclareMathAlphabet{\pazocal}{OMS}{zplm}{m}{n}

\usepackage{xspace}

\AtBeginDocument{%
  \providecommand\BibTeX{{%
    \normalfont B\kern-0.5em{\scshape i\kern-0.25em b}\kern-0.8em\TeX}}}

\usepackage{xspace}

\newcommand{\toolnamenox}{Guac}
\newcommand{\toolname}{\toolnamenox\xspace}
\newcommand{\salssa}{SalSSA\xspace}
\newcommand{\globalTool}{GlobalGuac\xspace}
\newcommand{\localTool}{LocalGuac\xspace}
\newcommand{\modelTool}{EnsembleGuac\xspace}
\newcommand{\exhaustTool}{ExhaustGuac\xspace}
\newcommand{\hyfm}{HyFM\xspace}
\newcommand{\fmsa}{FMSA\xspace}
\newcommand{\nw}{Needleman-Wunsch\xspace}

\newcommand{\accelmerger}{AccelMerger\xspace}

\renewcommand\footnotetextcopyrightpermission[1]{} 
\renewcommand\footnotetextcopyrightpermission[1]{}
\setcopyright{none}
\begin{document}

\title{Guac: Energy-Aware and SSA-Based Generation of Coarse-Grained Merged Accelerators from LLVM-IR}
\author{Iulian Brumar*}
\author{Rodrigo Rocha**}
\author{Alex Bernat*}
\author{Devashree Tripathy*}
\author{David Brooks*}
\author{Gu-Yeon Wei*}

    \affiliation{ 
      \institution{*Harvard University}
      \country{USA}
    }
    \affiliation{ 
      \institution{**University of Edinburgh}
      \country{UK}
    }




\begin{abstract}
  Designing accelerators for resource- and power-constrained applications
is a daunting task. High-level Synthesis (HLS) addresses these constraints through resource sharing, an optimization at the HLS binding stage that maps multiple operations to the same functional unit.

However, resource sharing is often limited to reusing instructions within a basic block.
Instead of searching globally for the best control and dataflow graphs (CDFGs) to combine, it is constrained by existing instruction mappings and schedules.

Coarse-grained function merging (CGFM) at the intermediate representation (IR) level can reuse control and dataflow patterns without dealing with the post-scheduling complexity of mapping operations onto functional units, wires, and registers. The merged functions produced by CGFM can be translated to RTL by HLS, yielding Coarse Grained Merged Accelerators (CGMAs).
CGMAs are especially profitable across applications with similar data- and control-flow patterns.
Prior work has used CGFM to generate CGMAs without regard for which CGFM algorithms best optimize area, power, and energy costs.

We propose \toolname, an energy-aware and SSA-based (static single assignment) CGMA generation methodology. \toolname implements a novel ensemble of cost models for efficient CGMA generation.
We also show that CGFM algorithms using SSA form to merge control- and dataflow graphs outperform prior non-SSA CGFM designs.
We demonstrate significant area, power, and energy savings with respect to the state of the art. In particular, \toolname more than doubles energy savings with respect to the closest related work while using a strong resource-sharing baseline. 
\end{abstract}




\makeatletter
\let\@authorsaddresses\@empty
\makeatother

\maketitle

\thispagestyle{empty}
\section{Introduction}
\label{sec:intro}

Manual accelerator designs attempt to reuse fine-grained control and dataflow application patterns through the design of processing elements (PEs). Such manual implementations can be found in cutting-edge domains such as deep learning and robotics~\cite{liu2015pudiannao,tambeLatest,suleiman2019navion}.
These multi-purpose accelerators greatly benefit from area and power optimizations due to the large number of impacted applications.
On the other hand, automated HLS-level resource sharing~\cite{lam2009rapid, brisk2004area} uses analytical models that assume a specific cost per operation. These works do not study the models' impact on the accelerator design's quality. The lack of model studies is due to the multi-layered nature of the HLS stages, including allocation, scheduling, and binding. Instruction cost models, when trained statistically, cannot minimize an objective function representing HLS allocation, scheduling, and binding to understand functional unit reuse profitability. Instead, a cost model mapping IR-level instruction characteristics to accelerator characteristics, e.g., power, performance, and area, is built and used as part of the HLS process when deciding which functional units to reuse. In the related work, not much attention has been paid to the mismatch between modeling instructions and modeling functional units. Models are rarely discussed at all in the context of efficient HLS resource sharing.

Resource sharing works at fine granularity, finding functional units without structural conflicts that inhibit reuse. Conflict-free reused chains of functional units accelerate small, semantically equivalent dataflow graphs scheduled for different cycles. In our experiments with state-of-the-art HLS and physical design tools, we find that this kind of resource sharing reduces energy consumption by 45.79\% with respect to accelerators not benefiting from resource sharing. (See Section~\ref{sec:resSharing} and Section~\ref{sec:resourceSharingResults})

Resource sharing, however, has a myopic view of what needs to be reused post-scheduling. It looks at accelerator reuse not from the perspective of dataflow, control flow, or input/output reuse but from a functional unit and structural conflict perspective. Some accelerator merging tools have sought to address this problem. ReconfAST~\cite{mokri2020early} produces merged accelerators at the loop level in a front-end compiler (clang), but at such a high level, optimization support is limited.
AccelMerger~\cite{Brumar2021accelmerger} leverages the function merging by sequence alignment (FMSA) algorithm developed in~\cite{rocha2019function}.
AccelMerger works at LLVM's intermediate representation (IR) level to provide coarse-grained function merging (CGFM), which merges functions and their call graphs. AccelMerger takes advantage of LLVM's rich framework for target-independent interprocedural optimization, rather than working on the abstract syntax tree.

These accelerator merging tools introduce a new way of designing accelerators that are more programmable than monolithic accelerators in that they can accelerate more than one top-level function or application. We call them CGMA, standing for coarse-grained merged accelerators. They are less programmable than finer-grained reconfigurable architectures, which spend significant area for functional unit (FU) and interconnect programmability. Finer-grained reconfigurable architectures suffer from significant communication overheads since they accelerate small control and dataflow patterns. A CGMA can approach the performance of the original monolithic accelerators from which it derives while consuming only a fraction of their total energy and area. CGFM can produce energy and area savings only for accelerators representing similar CDFGs. This similarity requirement puts pressure on the CGFM technique to better understand the impact of local merging decisions on end-to-end energy savings.

In this work, we develop CGFM models aimed at efficient CGMA design, and we identify the best CGFM algorithms for high CGMA energy savings. We choose energy savings as a proxy for co-optimizing power, performance, and area. An HLS tool can produce a monolithic accelerator starting from a function's high-level or IR-level representation. After selecting and merging a pair of related functions, we use HLS to produce a merged accelerator for the pair.

Few prior works can merge at such coarse granularities. To the best of our knowledge, the closest is \accelmerger~\cite{Brumar2021accelmerger}, which relies on the \fmsa IR algorithm for merging. That approach is limited in three key regards: 1) \accelmerger does not weight the instructions to be prioritized for merging in terms of their impact for area or energy savings, but instead applies a uniform weighting on the instruction types to be merged; 2) \accelmerger uses only one weighting of instructions per merge, which significantly limits profitability, since different linear models are more effective for different accelerator characteristics; 3) \accelmerger's non-SSA code representation sometimes leads to prohibitive overheads. Instead of SSA's phi instructions (mapped to multiplexers by HLS), it generates extra memory instructions, from which HLS produces expensive accelerator scratchpad accesses.

Prior work on CGMA does not analyze the potential of merging models to generate more area-efficient accelerator designs~\cite{Brumar2021accelmerger,mokri2020early}. Furthermore, optimization for energy efficiency has been outside the scope of prior CGMA research because of the long latency of HLS and physical design flows.

In this work, we propose \toolname, a novel compiler-based approach for merging accelerators using multiple non-uniform merging models.
We integrate \salssa~\cite{rocha2020effective}, a state-of-the-art CGFM approach, with our statistical cost model for energy usage. Our energy model is trained on post-physical-design estimates, improving its decision-making capabilities during the merge operation to achieve better energy savings.
In this study, we provide inter- and intra-application accelerator merges for a mix of 16 synthesizable applications from various domains, such as linear algebra, machine learning, and signal processing.
We also provide deeper post-physical-design analysis for power and device resource consumption, comparing the related work~\cite{Brumar2021accelmerger} with our novel solution.

\textbf{Overview of results that motivate the solution}. When applied to an FPGA target using a strong resource-sharing baseline, \toolname reduces LUTs by 7.04\%, Flip-Flops (FF) by 11.17\%, and energy consumption by 17.08\% on average. By contrast, the closest related work only improves energy savings by 7.38\%, LUTs by 1.06\%, and FFs by 2.01\%. We demonstrate these results with high accuracy over 16 applications in an LLVM-to-bitstream setup using a dataset of 10986 merged accelerators.

Our contributions are summarized as follows:

\begin{itemize}
    \item {\bf Energy-Aware CGFM for CGMAs.} We implement a two-step energy-saving strategy. First, we use arithmetic-, control- and memory-centric CGFM models to address different bottlenecks in the merged function IR. Second, we select the final merged function version through an ensemble model that chooses the most profitable function for acceleration based on LLVM instruction savings. Related works on coarse-grained and fine-grained accelerator merging abstract away later synthesis stages, such as physical design, and rely on HLS-level quality of result estimates. As highlighted in~\cite{modelfccm}, HLS-level estimates can be highly unreliable even for area, with 125\% LUT and 98\% Flip-Flop mean relative error (MRE). To create energy-efficient merged accelerators with high confidence in the quality of results, \toolname provides an automated tool flow from high-level C/C++ to post-place-and-route and bitstream generation.
    \item {\bf Evaluation for Both Inter- and Intra-Application Merging.} We perform an exhaustive evaluation of possible merged accelerator candidates for the MachSuite applications~\cite{reagen2014machsuite}, originally created for accelerator design. We observe that HLS and physical design tools optimize certain application patterns in a subset of applications. Still, we demonstrate a more general approach to accelerator merging at a coarse granularity. 
    \item {\bf Leveraging Static Single Assignment (SSA) for Accelerator Merging.} CGFM has been used for accelerator merging in prior work~\cite{Brumar2021accelmerger}, but without the benefit of SSA form~\cite{lattner2004llvm} throughout code generation. That increases multiplexing costs, both for logic and for energy. In this paper, we show that just using the latest CGFM techniques out-of-the-box to achieve higher energy savings is not feasible unless those techniques optimize globally. We show that even though new local CGFM techniques have potential to reduce control logic, global techniques deliver energy savings of over 15.91\% compared to less than 5\% improvement for both local techniques and non-SSA approaches.
\end{itemize}


\section{Motivation}

\label{sec:motivation}

\begin{figure*}
  \centering
  \includegraphics[width=0.9\linewidth, trim={0cm 0cm 0cm 1cm}]
	{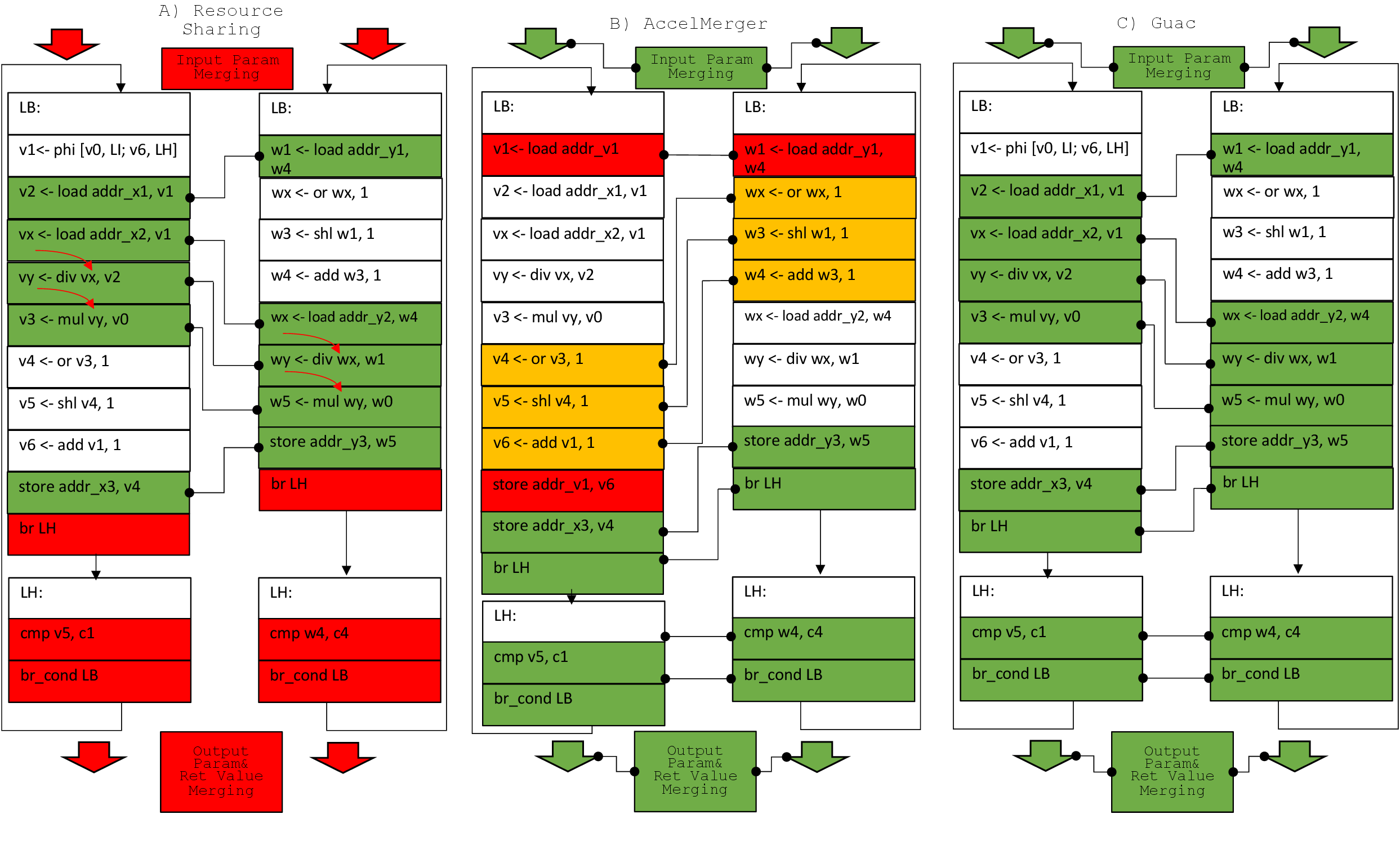}
  \caption{Resource Sharing, AccelMerger and \toolname LLVM-IR level example.}
  \label{fig:resourceSharAndAccelMCode}
  \vspace{-3.7mm}
\end{figure*}%

The state-of-the-art in area and energy reduction techniques is resource sharing, binding time, and HLS optimization. Figure~\ref{fig:resourceSharAndAccelMCode} illustrates resource sharing on the left-hand side. This figure illustrates the alignment of two functions at the LLVM IR level to reduce area and power consumption by merging those functions. We represent operation reuse with a connection between the multiply, divide, and memory operations. Resource sharing avoids reusing smaller functional units like those corresponding to additions or bitwise operations since the cost of multiplexing and routing the operands can be too high.

In this example, selecting between register 'vy' and 'wy' for the first multiplication operand and between 'v0' and 'w0' for the second one requires two multiplexers. Reuse at coarser granularities often removes the need for such multiplexers and extra operand routing since reuse is implemented at the level of the control and data flow graph (CDFG). This issue is represented in Figure~\ref{fig:resourceSharAndAccelMCode} with red arrows between the definitions and uses of some registers, which is easy to identify but less obvious at binding time, post-scheduling once we have two mapped loops. Furthermore, the fact that resource sharing does not merge branch and comparison operations results in larger control circuitry and ultimately interacts negatively with dataflow sharing across high-area instructions. Similarly, resource sharing misses an important opportunity to merge the inputs and outputs for the circuits it generates. Co-optimizing the reuse of input and output wiring with the reuse of the CDFGs is more appropriate for CGFM-related techniques, including \toolname's. In fact, functional unit (FU) merging in resource sharing occurs at a lower level of abstraction. We only illustrate the result at the IR level for one-to-one comparison with CGFM-related merging.

Resource sharing is designed for FU reuse within the scope of a function, not for cross-function FU sharing. To establish a strong baseline for the experiments in coarser-grained merging described later, we enable resource sharing between two functions by creating a new top-level function containing both function bodies. (See Sections~\ref{sec:resSharing} and~\ref{sec:concat}.)

We represent the closest related work in CGMA generation, namely \accelmerger, in the middle of Figure~\ref{fig:resourceSharAndAccelMCode}. \accelmerger manages to merge some of the instructions whose sharing is most beneficial. These include a store, a branch operation, and a comparison operation, all highlighted in green. Such reuse reduces accelerator scratchpad access logic, control and scheduling logic, and dataflow-related logic, like that between the cmp operation and the br\_cond instruction.

CGMA-based techniques such as \accelmerger can significantly improve inter-function FU reuse, including dataflow and control flow between instructions. The key insight is that even though the number of possible IR alignments between two input functions is exponential, due to their control and dataflow graph representation, linearizing them and then aligning optimally yields high area savings in practical time. Furthermore, since coarse-grained merging approaches like \accelmerger and \toolname merge at the IR level, feedback about circuitry reuse can be provided to the designer in the earliest stages of accelerator design. This is unlike resource sharing, which requires the application to be in a synthesizable format (e.g., with simple array data structures and side-effect-free functions).

A major drawback of \accelmerger is that the mul operation is not reused, despite its high area and energy costs. Instead, \accelmerger decides to merge the bitwise or, the shift (shr) and the add operations, highlighted in yellow, because it lacks an energy-driven model. Creating accurate energy models for this type of function alignment is challenging since operations are not directly mapped to FUs by HLS. Instead, this mapping is the result of a three-step HLS process: allocation, scheduling, and binding. Furthermore, these steps depend on the device to which these functions are lowered, which determines the technology used, the number and type of hardened FUs, and their connection topology. Section~\ref{sec:fmsa} describes in more detail the challenges of achieving alignments that result in high energy savings.

\accelmerger merges functions in a non-SSA format, in which loads and stores replace phi instructions. Phi instructions are critical for handling control flow but have been hard to merge from a code generation perspective~\cite{rocha2020effective}. Figure~\ref{fig:resourceSharAndAccelMCode} illustrates an alignment that is undesirable due to \accelmerger's choice of CGFM algorithm. The phi instruction highlighted in red is implemented by a load and a store in non-SSA format, and the load will be merged by AccelMerger against the load in the right-hand-side function. At the code generation stage, AccelMerger will attempt to map the merged function code back to the SSA form but will fail to match one store and one load back to a phi instruction because the merged load did not belong to a phi instruction and, therefore does not have a corresponding phi-related store. This prevents HLS from finding the optimal synthesis: one multiplexer corresponding to the phi instruction and one memory-reading FU corresponding to the load. Instead, it generates one load FU and one store FU, due to the demotion of the phi instruction.

 \toolname addresses the CGFM-based CGMA generation problems at the IR level by implementing a two-stage merging cost model (ensemble model). The first stage applies multiple instruction-level alignment models and produces the merged function IRs for all the models. The second stage evaluates these IRs with the help of a coarse-granularity function-level statistical model that evaluates the energy consumption of all the merges. This enables the multiply instruction alignment in Figure~\ref{fig:resourceSharAndAccelMCode} instead of the lower cost shift and addition operations. 

We also show that cutting-edge local merging algorithms that improve control logic reuse are not as useful as global algorithms that merge across the boundaries of basic blocks. By evaluating multiple code generation and alignment algorithms, \toolname improves phi instruction handling and (more importantly) increases overall merged accelerator energy savings. 

\section{Background}

\label{sec:background}

\subsection{High Level Synthesis}

High Level Synthesis translates high-level code like C/C++ to Verilog or VHDL. Its stages typically include Allocation, Scheduling, Binding and RTL generation~\cite{hlsintroduction,soda,llvm4hls}. We focus on the first three of these since the last stage just translates from the HLS internal representation of functional units and registers to RTL.

\textbf{Allocation} uses device operation modeling to produce an initial estimate of the area required by an accelerator, especially in terms of functional units.

\textbf{Scheduling} is the phase in which the HLS tool distributes operations across cycles, taking into account their dependencies. Dependent instructions are chained together within the same cycle if they do not violate a target cycle time. In this step the Control Logic of the accelerator is generated, typically as a finite state machine. Neither allocation nor scheduling has a concept of functional units on which operations will be scheduled.

\textbf{Binding} constitutes the mapping of operations onto functional units. In this step, HLS executes the resource sharing optimization by mapping multiple operations onto the same functional unit if they are not executed in the same cycle. If the dataflow style of the reused functional unit chain is identical to a previously bound operation chain, no multiplexing is needed. However, input and output operands usually need to be routed differently, and expensive multiplexers need to be used. For this reason, only long latency instructions or chains thereof are reused.

\subsection{Resource Sharing}

\label{sec:resSharing}

Figure~\ref{fig:accelmergingOverview}A represents resource sharing, which shares high-area functional units among multiple operations. The HLS resource sharing optimization introduces multiplexing logic. In this figure, the shared functional units are highlighted in green. The non-shared circuit elements are shown in white and red. The red ones are suboptimal merging outcomes, due to the limitations of the resource sharing technique. In white, we represent circuit elements that are not merged, either because of mismatches between functions or the high cost of multiplexing. 

Next, we describe how we set up a strong baseline for resource sharing.


\subsection{Concat-Style Accelerator Merging}

\label{sec:concat}

Resource sharing operates within the scope of a single top-level function, while CGFM operates on a pair of top-level functions. The question is how to fairly set up resource sharing to compare it with CGFM. That requires enabling resource sharing not only within the two functions intended for merging, but also across their boundaries.

To fairly evaluate the potential of resource sharing when merging two functions $f_1$ and $f_2$, we expose both functions to the resource sharing optimization. We concatenate the high-level code and parameters of $f_1$ and $f_2$ at the IR level. 

\begin{figure}[t]
  \centering
  \vspace*{-22.5em}
  \includegraphics[width=0.95\linewidth, trim={0cm 0cm 0cm 4cm}]{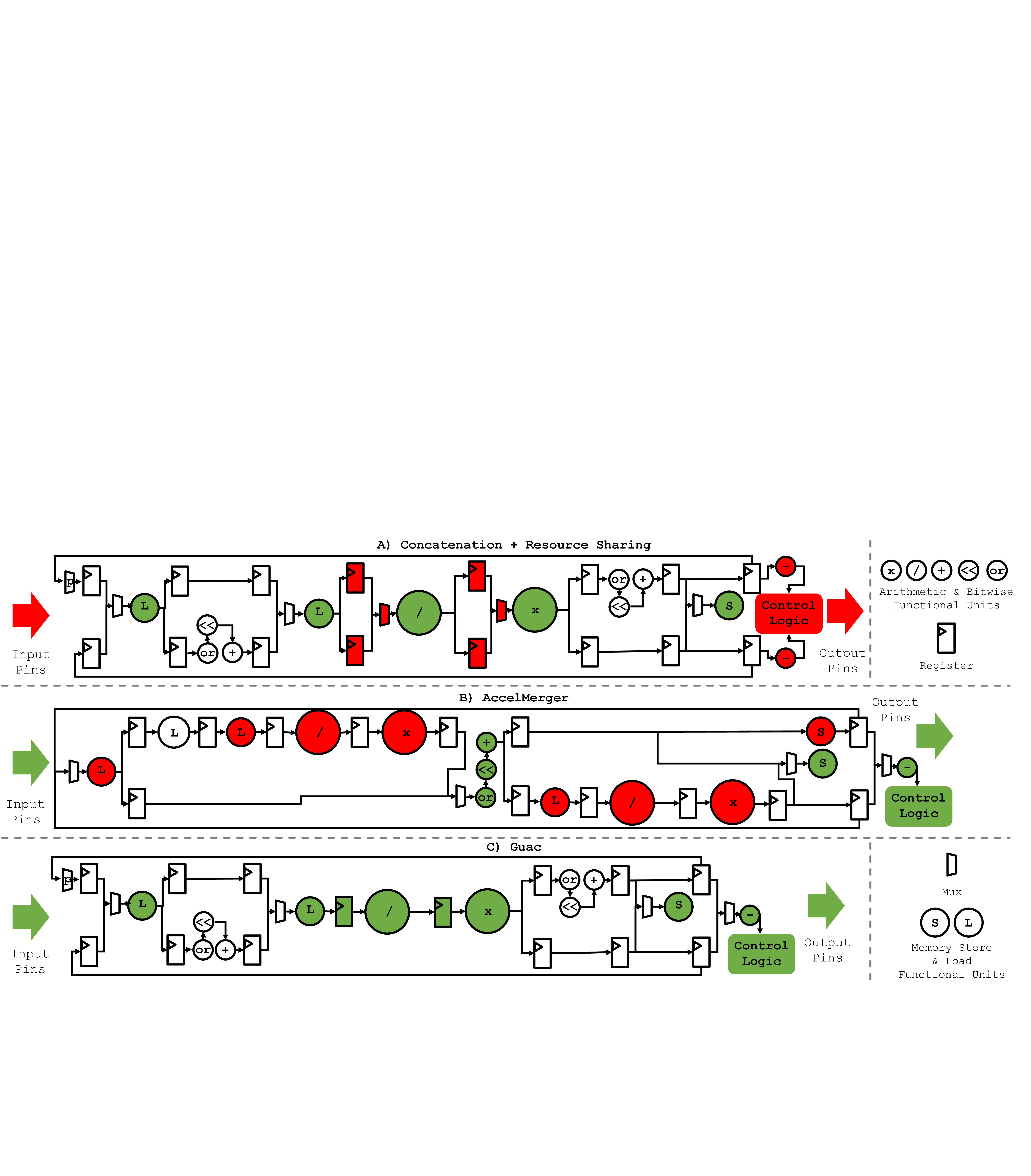}
  \vspace*{-8.5em}
  \caption{Schematics illustrating the merging circuit-level benefits: A) Function concatenation followed by HLS with resource sharing, B) Coarse-grained merging using \accelmerger. We indicate in red the alignment between a demoted phi instruction and a load as well as the resulting overhead store, C) Coarse-grained merging via \toolname. The wiring across the three techniques is simplified. The green register in the case of C) indicates higher savings when operands have matching dataflow across the merged functions.
  }
  \label{fig:accelmergingOverview}
\end{figure}

We call this strategy Concat or Concatenation. Moreano et al.~\cite{moreano2005efficient} manually implement a simplified version of Concat that works for single basic block RTL modules with easy-to-concatenate input/output parameters. Concat has been automated for arbitrary control and dataflows at the level of (pre-HLS) LLVM IR~\cite{Brumar2021accelmerger}. Figure~\ref{fig:accelmergingOverview}A indicates in red the lack of reuse of control logic due to non-merged branch and comparison instructions. It also highlights in red the lack of dataflow reuse leading to extra registers and multiplexers. Finally circuit input/output pins are not reused. HLS trusts the programmer to design optimal input and output interfaces. However this type of manual merging can be tedious and error prone for coarse-grained merging, even though input and output merging represents an opportunity for resource savings. We will later see that this is a great opportunity for automation (missed by resource sharing) when the circuit inputs and outputs have matching types and the circuitry they connect to has high similarity.  

To measure savings in this baseline configuration, we define the cost of mapping functions $f_i$ and $f_j$ to monolithic accelerators (resource sharing only locally within the CDFGs of the functions but no functional unit reuse across the borders of $f_i$ and $f_j$) as $C_{monolithic, Y}\left(f_i\right)$ and $C_{monolithic, Y}\left(f_j\right)$. By contrast, mapping $f_i$ and $f_j$ to a merged accelerator via concatenation costs $C_{concat, Y}(f_i, f_j)$. Here $Y$ represents both the area  (LUTs, FFs, DSPs) and the energy dimensions. The improvements in a given dimension $Y$ are $I_{Y}(f) = \frac{C_{monolithic, Y}\left(f_i\right) + C_{monolithic, Y}\left(f_j\right)}{C_{concat, Y}\left(f_i, f_j\right)}$.



We measure latency overheads by using the merged accelerator to execute each of the functions that it implements. We define the latency overhead of a function when run on a merged accelerator as $L_{ovh}\left(f_i\right) = \frac{L_{concat}\left(f_i, f_j\right)}{L_{monolithic}\left(f_i\right)}$, with $L_{concat}\left(f_i, f_j\right)$ being the latency for running $f_i$ on the merged accelerator produced by synthesizing the concatenation of $f_i$ and $f_j$, and $L_{monolithic}(f)$ being the latency for running $f$ on a dedicated accelerator that was generated by HLS directly from $f$.  We compute the merged accelerator latency generated from merging functions $f_i$ and $f_j$ by averaging the overhead of running $f_i$ and $f_j$ on the merged accelerator: $\frac{L_{ovh}\left(f_i\right)+L_{ovh}\left(f_j\right)}{2}$.  In our results we report the percentage of both energy and area improvements as well as latency overheads by computing $\left(I_{Y}\left(f\right) - 1\right)\times100$ and $\left(L_{ovh}\left(f\right) - 1\right) \times 100$.

\subsection{Function Merging by Sequence Alignment} \label{sec:fmsa}


\begin{figure}
  \centering
  \includegraphics[width=0.6\linewidth]{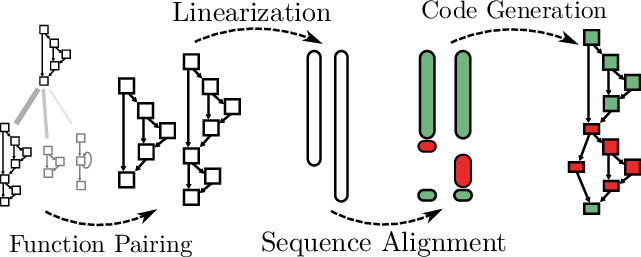}
  \caption{Overview of coarse grained function merging workflow. Reproduced with the author's permission~\cite{rocha2019function}.}
   \label{fig:fmsa-overview}
  \vspace*{-1.5em}
\end{figure}%

\textbf{\fmsa overview}. \fmsa~\cite{rocha2019function} was the first CGFM algorithm able to merge arbitrary non-identical pairs of functions introduced with the goal of reducing code size. It was later used in the context of high level synthesis~\cite{Brumar2021accelmerger} for area reduction (\accelmerger). We now focus on the \fmsa flow at the LLVM IR level.
The algorithm consists of 
three major stages: choosing which functions to merge, producing the merged function, and estimating the merging profitability.


\textbf{Function Pairing}. The first step in function merging is "Function Pairing" as indicated in Figure~\ref{fig:fmsa-overview}. In this paper, we perform merging across multiple applications and we merge all pairs of functions in the input applications since some pairings might be profitable only for some techniques and not for others.

\textbf{Function Linearization}. For a given pair of functions, merging them requires identifying similar code segments in the two functions that can profitably be merged.
To this end, \fmsa~\cite{rocha2019function} first linearizes the input functions, representing the CFG as a sequence of instructions.

\textbf{Alignment}. Then the \nw (NW) sequence alignment algorithm identifies regions of similarity between the two functions.
The alignment algorithm is applied to the linearized sequences of the whole input functions. The NW algorithm allows weighting each operation. In this work, our alignment models provide the NW weighting of instructions. 
Finally, the resulting alignment is used to generate the merged function.

\textbf{Code Generation}. During code generation, the list of parameters from both functions is also merged using a greedy algorithm that tries to merge as many parameters as possible.
Parameters of the same type are shared between the functions, without necessarily keeping their original order. 
The merged list of parameters also includes an extra function identifier. All the calling sites for the input functions are also modified to point to the newly merged functions with the right parameter order and original function identifier. 

\textbf{New Basic Blocks for Contiguous Aligned and Unaligned Instructions}. Once the function definition has been created, the control-flow graph of the merged function is produced by traversing both the input functions and the aligned sequence.
The merged control-flow graph uses the function identifier for control flow divergence where the two functions differ, but converges to the same basic block where they are similar.
Operand selection is also introduced where two merged instructions have different operands.
Finally, \fmsa applies an SSA reconstruction algorithm on the merged function.

We believe CGFM-based CGMA generation tools should leverage the flow from alignment through code generation to apply different alignment models, producing different versions of the merged function. Owing to the nature of sequence alignment algorithms such as NW, these alignment models are linear. However, after code generation we have the freedom to evaluate the quality of the IR with more sophisticated, non-linear models, able to extract more information about accelerator power and area characteristics by analyzing function-level characteristics such as number of instruction counts for each instruction type. The potential for such alignment and code generation models is explored for the first time in \toolname.



\textbf{SSA-Level Limitations}. A major limitation in \fmsa is its inability to handle phi nodes, a core feature of the SSA form.
In order to work around this limitation, \fmsa performs register demotion (\verb|register-to-memory|) in all functions prior to linearization, replacing all phi nodes with a stack-based store-load mechanism.
This process tends to create larger functions, however, lowering the quality of the function merging optimization, as stores and loads corresponding to phi instructions can be merged against non-phi-related loads and stores. Extra unaligned loads and stores consequently add a significant overhead when mapping those instructions to functional units at the allocation step of HLS.
We discuss in Section~\ref{sec:salssa} how to address this limitation, and we show in Section~\ref{sec:irModelling} why merging phi nodes and control-flow-related instructions is relevant for accelerator merging.

\textbf{CGFM for CGMAs}. \accelmerger~\cite{Brumar2021accelmerger} uses \fmsa as part of infrastructures for merged accelerator design based on abstract, post HLS LUT estimates, without considering \fmsa's potential for power reduction, nor breaking down the benefits of \fmsa when applying it to reduce different resource types or power consumption. As indicated in Figure~\ref{fig:accelmergingOverview}B, with \accelmerger, full data and control paths can be reused, including registers and functional units. This reuse across the CDFG allows reducing the number of multiplexers when there is a match in the control flow. Some of the savings in CDFG reuse, however, are hindered due to overhead memory operations that \fmsa cannot transform back into phi instructions, since it performs merging in the non-SSA format, and when functions are aligned, the components of phi instructions get aligned with other instructions. We highlight this issue in Figure~\ref{fig:accelmergingOverview}B with the red (aligned but not profitable in this case) load (L) unit. Even though this is the result of aligning the load instruction against the phi instruction, it results in extra store functional units later in the design of the top function.

\subsection{\salssa and \hyfm} \label{sec:salssa}

\salssa~\cite{rocha2020effective} is a CGFM algorithm focused on improving \fmsa code generation by supporting the SSA representation. It avoids the code bloating problem introduced by register demotion and increases the chances of generating profitable merged functions. \salssa's code generator achieves this by treating phi nodes as special instructions, excluding them from function alignment since in most cases \fmsa would miss-align the equivalent alloca, store, and load operations. This leads in practice to more saved instructions. Theoretically, there are cases when \fmsa will merge the alloca, load and store instructions corresponding to the phi instructions correctly against other phi-related alloca, loads and stores. \salssa tends to save more because such perfect phi matches in non-SSA format are unlikely, and even not merging phi instructions at all is more easily mapped onto multiplexers at synthesis time. 


While \salssa, like \fmsa, tries to find an optimal alignment globally across all the instructions of the merged functions, \hyfm~\cite{rocha21hyfm} emphasizes local basic block merges. \hyfm first tries to find the best match of basic blocks according to static instruction count similarity (fingerprint comparison) and then merges instructions within basic blocks. The advantage of this technique is not only that it reduces sequence alignment execution time, but also that it is guaranteed to reduce the number of branches in the function's control flow, since every basic block ends with a branch. This reduces the number of predicated instruction executions in the equivalent accelerators and control-related logic. The disadvantage of this approach is that merges are not global across all basic blocks, so an opportunity is missed to maximize function-wide area or power savings.



\section{\toolname}

\label{sec:fmerging}

\begin{table}[]
\begin{tabular}{ccccccc}
\hline
 & \textbf{Load} & \textbf{Store} & \textbf{Phi}   & \textbf{Br}    & \textbf{Alloca} & \textbf{Select} \\ \hline
\textbf{LUTs}       & 1:19 & 1:37  & 1:15  & 1:133 & 1:120  & 1:65   \\ \hline
\textbf{FFs}        & 1:21 & 1:16  & 1:19  & 1:51  & 1:0    & 1:64   \\ \hline
\textbf{DSPs}       & 44:1 & 1:0   & 197:1 & 6:1   & 1:0    & 4:1    \\ \hline
\end{tabular}

\caption{Weight of IR/LLVM control-flow related instructions when mapping them onto FPGA resources. The first number in each cell refers to the number of static instructions. The second number following ':' represents the number number of resources in the row required to synthesize that static instruction. These numbers result from a multilinear regression analysis of the MachSuite applications.}
\vspace*{-0.1em}
\label{table:instrCost}
\end{table}

\subsection{CGFM for accelerator energy efficiency}

\label{sec:irModelling}

\textbf{Alignment and post-code-generation energy model}. The main shortcoming in the related work is that CGFM techniques do not merge for energy savings. There are two possible approaches to making CGFM energy-savings-aware. One way is by using alignment models. Algorithms like Needleman-Wunsch (NW) can align sequences with a linear model, attributing a cost to each instruction type. The result is that NW finds the alignment with the lowest cost due to the algorithm's optimality guarantees. 

 The second possible energy model considered in this work is post-code-generation. It is applied on the resulting CDFG after aligning two functions. Post-code-generation models evaluate the potential energy savings directly using the features of the merged function instead of some abstract representation of the function (e.g., the sequences used in the sequence alignment algorithm).
 
 We use alignment-time merging to create multiple versions of a merged function, while taking account of the accelerator back end. Then we apply a code-generation-time statistical model based on random forests to select the accelerator version most likely to generate energy savings.


\textbf{Control, memory, and arithmetic alignment models}. Alignment models are analytical, and we use them to emphasize saving either control, memory, or arithmetic instructions. The intuition for this type of model is that array-style architectures, including FPGAs, have limited resources per slice (e.g.,  multiplexers, flip-flops, lookup tables, or DSPs). The per-slice logic is under-utilized when an application is \mbox{memory-}, compute- or control-intense. Still, a good balance of resource usage allows HLS to map more operations per slice, reducing routing and multiplexing costs. In the three alignment-time models, we put a higher weight (twice that on the other instructions) on either of these instruction types. 

These savings are represented in Figure~\ref{fig:accelmergingOverview}C in the proper merging prioritization of the high-cost division, multiplication, and memory instructions. The difficulty here is that modeling needs to operate from a higher level of abstraction than resource sharing from the IR level. Resource sharing benefits from accurate knowledge about the mappings between operations and functional units, whereas at the IR level, we need to use both per-operation models to guide the merging process, as well as overall evaluations of the generated CDFG in the IR (post-code-generation models).

For the control models, we prioritize the merging of all the LLVM branch instructions (conditional, unconditional, direct, and indirect), phi instructions, compare instructions (cmp and fcmp), and the switch statement. For the memory model we include load, store, alloca, fence, extractelement, extractvalue, getelemptr, insertvalue, insertelement, and shufflevector. For the arithmetic model, we include all floating point, integer, boolean, bitwise and casting operations. The reason for using these specific models is the pressure each type of instruction puts on the FPGA slices. Implementing CGMAs for a target other than FPGA, such as ASIC, will benefit from further fine-tuning of the CGFM model to the cell library characteristics.


\textbf{The need for post-code-generation models}. In our experiments and in related work on guiding IR-time optimizations for accelerator back ends~\cite{fpgaarchimpactonrs,llvm4hls}, we find that no single alignment model can improve merges across applications. There are three reasons why this is so.  First, different application mixtures create different bottlenecks.  Second, alignment algorithms rely on linear models to determine the weight of an operation whereas the process of mapping instructions to hardware is highly non-linear.  Third, many overhead instructions, such as extra selects for the operands of merged instructions, are not known at alignment time, but are available after code generation.

\toolname's post-code-generation random forest uses the number of saved instructions of each instruction type, compared to the original functions, to predict energy savings. This prediction is performed on the three versions of the merged function with either control, memory, or arithmetic focus. We collect a dataset of true energy savings (target variable) on a thousand post-place-and-route merged accelerators and their operation-level savings (features) to train the random forest model.

\begin{figure*}
  \centering
  \includegraphics[width=0.9\linewidth, trim={0cm 0cm 0cm 0cm}]
	{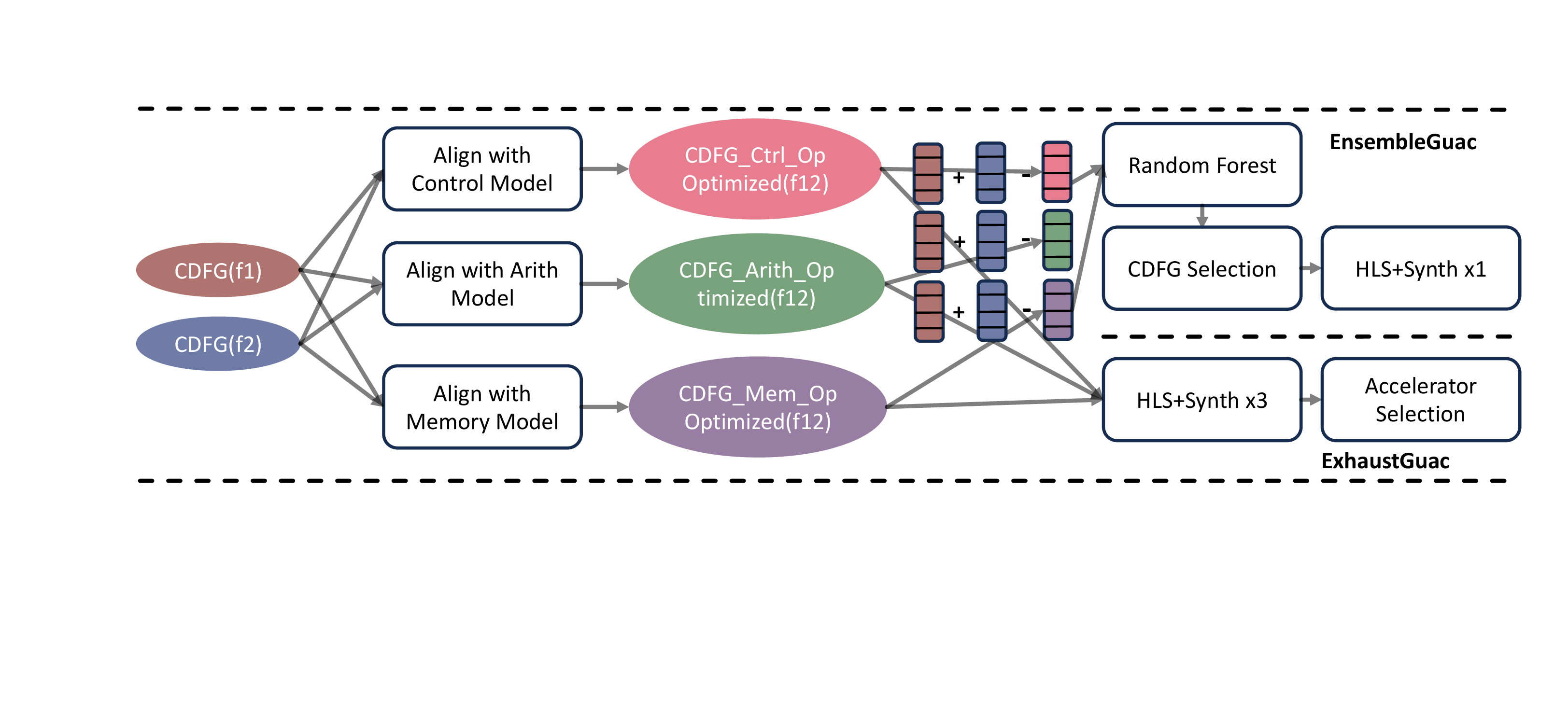}
  \vspace{-20.1mm}
  \caption{Alignment-model based \toolname configurations \modelTool and \exhaustTool.}
  \label{fig:alignmentAndPostCodegenModel}
  \vspace{-1.1mm}
\end{figure*}

\textbf{\modelTool vs \exhaustTool}. Figure~\ref{fig:alignmentAndPostCodegenModel} illustrates our two-step model in a configuration used in this paper named \modelTool. We also present a configuration called \exhaustTool, which uses only alignment models but relies on HLS to evaluate the final quality of the merges with different alignment models. We can see the flow from two input functions $f_1$ and $f_2$, and their respective CDFGs represented with elliptical shapes. These functions are merged using the three alignment models, thanks to the NW algorithm cost model employed in all the CGFM approaches discussed in this paper. This algorithm allows one type of instruction to be prioritized by giving it a higher cost than the other types.


The result is three CDFGs that emphasize the optimization of control, arithmetic, or memory instructions. Next, we extract instruction counts from both the input functions and the merged ones, and we produce an LLVM IR saved instructions vector corresponding to each of the alignments. A random forest model uses these vectors to predict which alignment model will save more energy for this particular merged function. The rationale is that we use savings in a quickly retrievable quantity (IR counts of different merges) to predict energy savings, then apply time-consuming HLS and synthesis only to the most promising CDFG. We call this configuration \modelTool since it uses the random forest to select among an ensemble of alignment models. Alternatively, \exhaustTool runs HLS and synthesis on all three CDFGs (incurring a large time penalty) and then selects the accelerator with the lowest energy consumption. This configuration is more time-consuming but it will select the most efficient version of the merged accelerator whereas \modelTool can occasionaly miss the most energy efficient configuration.

The random forest we use for post-code-generation in \modelTool is trained on a thousand merged accelerators among the 10986 possible merges among the functions in the MachSuite applications using scikit-learn. The rest of the points are used for testing. To find a well-fitting model we use cross-validation and grid search over the random forest maximum depth, number of estimators, and maximum features. The hyperparameter space we explore contains ten equidistant values between 1 and 30 for maximum depth (depth of each tree), ten equidistant values between 10 and 250 for the number of estimators (number of trees in the random forest) and three possible values for the maximum number of features to use for prediction among the LLVM instruction types. For the maximum number of features we use 12.5\%, 25\%, and 50\% of the 64 LLVM instruction types. The hyperparameter space contains a total of 300 values. On this particular problem we found Random Forests to perform better than a variety of other models including the multilayer perceptron.

\subsection{CGFM merging technique selection for CGMAs}

Choosing efficient CGFM techniques for CGMA design requires knowing the cost of each IR static instruction in terms of allocated hardware resources. We perform a statistical analysis of all merged accelerators across all applications to inform us about resource usage of instruction types. Specifically, we use a multiple linear regression model per target variable (LUTs, FFs, and DSPs). This model assigns a weight to each instruction type, resulting in the data in Table~\ref{table:instrCost}.

Table~\ref{table:instrCost} represents the cost of key instructions used by HLS flows to generate an accelerator's control, data, and memory access logic. The content of each cell represents the number of resources required to implement an operation on average. 

For example, in terms of LUT consumption, branch instructions (Br) have the most impact, resulting in an average of 133 LUTs allocated per branch instruction. Since DSPs are coarser-granularity resources than LUTs and FFs, one DSP is observed every so many instructions of each type. For example, for every 44 synthesized load instructions we observe one allocated DSP resource.

The modeling in Table~\ref{table:instrCost} shows that branch instructions have the most impact on  reducing LUTs and the second most important impact on FFs and DSPs. Together with their high area footprint, this motivates using function merging techniques that prioritize merging branches.
By merging basic blocks one-to-one, \hyfm also maximizes the chance of merging branch instructions, since each basic block terminates with a branch in LLVM.

\textbf{\accelmerger feasibility}. Also, as described in Section~\ref{sec:background}, \fmsa-based accelerator merging~\cite{Brumar2021accelmerger} relies on alloca, load, and store instructions to emulate the behavior of phi instructions. \fmsa relies on the \texttt{register-to-memory} LLVM pass to map low-cost phi instructions (correlated with 15 LUTs, 19 FFs, and $\frac{1}{197}$ DSPs) to a load, an alloca, and stores for each operand. The load is placed at the site of the phi instruction and the alloca in a basic block that dominates the phi-related load and store (the phi-related stores and load need to occur at an allocated memory location). \texttt{register-to-memory} creates a store at the site of each assignment of a register that is used by the phi instruction. There is no guarantee that \fmsa will merge these load, store, and alloca instructions against other phi-related instructions. An \fmsa-based phi misalignment, therefore, has an overhead of at least the 120 LUTs corresponding to an alloca, $37\times2 = 74$ LUTs due to at least two incoming stores per phi instruction and 19 LUTs per load in addition to all the FFs and DSPs corresponding to these instructions as shown in Table~\ref{table:instrCost}. 

Whenever any of the phi-related instructions are merged against non-phi-related instructions, the \texttt{memo-to-reg} pass will not recover the low-cost phi instruction. In this paper, we analyze SSA-based CGFM techniques that avoid high area and energy penalties.   Based on this analysis, as well as our experimental results, we show that SSA-based merging algorithms are most profitable for accelerator merging. This is reflected in Figure~\ref{fig:accelmergingOverview}C since the phi instruction is left intact, enabling the reuse of the first load of the top circuit (first function). All the benefits present in \accelmerger in terms of control circuitry as well as input/output merging are preserved, since the merging algorithms with and without SSA are identical from that perspective.



\textbf{Global, SSA-based merging feasibility}. The first novel CGFM technique we evaluate for this paper is \salssa, which performs global merging like \fmsa, but operates directly on the SSA representation of the code.
Compared with \fmsa, \salssa reduces the number of memory loads by $9.41\%$, stores by 22.74\%, and alloca instructions by$19.55\%$. This results in less circuitry related to memory operations as well as reductions in accelerator wiring.
Even though \salssa tends to produce over $2\times$ more phi nodes than \fmsa, most can be eliminated by register coalescing strategies~\cite{chaitin1981}. Unlike select instructions, phi nodes are not implemented by multiplexers. 


\textbf{Local, SSA-based merging feasibility}. The main problem with global merging techniques is that they do not emphasize control-heavy instructions and multiplexer reduction in high-control-flow scenarios. We evaluate \hyfm, a local and newer, merging technique than \salssa, for its potential to reduce control-flow-related bottlenecks. As shown in Table~\ref{table:instrCost}, merging branch instructions can have a dramatic impact on reducing LUT and FF resource utilization (133 and 51 units, respectively, per branch).

As we will see empirically in Section~\ref{sec:results}, even though some applications do indeed benefit more from \hyfm than from \fmsa for energy reduction, global methods benefit CGMA generation much more on average. As a global and SSA-friendly technique, \salssa delivers the highest benefits overall. For this reason, all our \toolname configurations use \salssa as the main merging technique.

Prior work on coarse-grained merging has focused on intra-application merging, i.e., trying to merge accelerators corresponding to functions in a single application~\cite{Brumar2021accelmerger}. We evaluate \toolname on both inter- and intra-application scenarios.

\begin{figure*}

  \centering
  \includegraphics[page=2,width=0.95\linewidth,
		trim={0cm 0cm 1cm 1cm}]
		{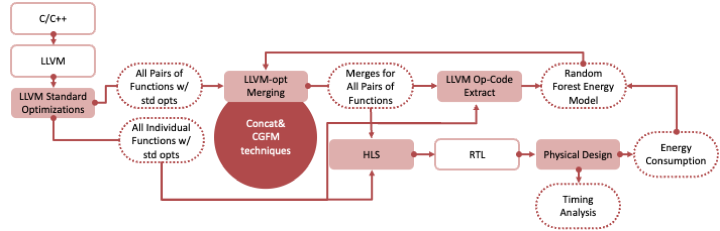}
  \caption{End-to-end flow for \toolname, \accelmerger and concatenation+resource sharing evaluation.}
  \label{fig:endToEndFlow}

\vspace*{-1.6em}
\end{figure*}%

\section{Implementation}
In this section, we describe our end-to-end solution.
Figure~\ref{fig:endToEndFlow} shows the process by which we extract merged accelerators from the high-level source code, how merged functions are being produced and how we map both merged and non-merged functions to accelerators.

The flow for merging consists of a front end that collects important information about the application, including generating the testbench for the cosimulated functions. Then the middle end step merges functions with the goal of achieving large area savings while trading off the minimum amount of performance. The back end converts both merged and non-merged functions to RTL via HLS. Then it performs technology mapping and place-and-route for accurate FPGA area and power data collection.

Our experiments measure area savings, latency overhead, and energy savings with respect to fine-grained HLS resource sharing. Specifically we measure CGMA improvement for two functions $f_1$ and $f_2$ on resource $Y$ using the formula $I_{CGMA, Y} = \frac{C_{concat, Y}(f_1, f_2)}{C_{CGMA, Y}(f_1, f_2)}$. Since we are directly comparing the merged accelerators, we measure latency improvement instead of overhead since the HLS layer sometimes trades area savings for latency and power improvements.

The main focus in this paper is on improving CGMA energy efficiency. We focus on the energy component of the used resources. The selection of a device that is a good match for a given application, so that the dynamic component of energy dominates the design, is outside the scope of this work.

\subsection{Standard Optimizations and Co-simulation}

Source code is compiled down to LLVM IR. Standard optimizations among \verb|-O1|,\verb|-O2| and \verb|-O3| are applied on the compiled IR.  In our experiments, we find that \verb|-O1| creates an IR that produces the lowest area and power consumption across the \verb|-Ox| configurations, allowing the HLS tools to fine tune code for FPGA targets. All the reported merged accelerators, including baseline concatenation-based accelerators, are produced with \verb|-O1| on the result of the merged function as well as on the input functions.

Our HLS tool requires a \texttt{.xml} file to specify the input values to use for the co-simulated functions. Given the large number of merged functions to evaluate, we automate the generation of this \texttt{.xml} file by creating an LLVM instrumentation pass and a run-time library.

\subsection{CGFM techniques and Resource Sharing baseline}
\label{sec:configs}

We label different LLVM-to-bitstream flows and merging techniques so that we can compare their PPA efficiency: 
\begin{itemize}
\item \accelmerger for \fmsa-based flow to represent the limitations of the related work in generating efficient designs in terms of models and SSA-based merging.
\item \localTool for \hyfm-based flow to evaluate local merging for its potential in reducing control-related logic and for its SSA benefits.
\item \globalTool for \salssa-based flow to evaluate SSA-based, global merges.
\item \modelTool for \salssa plus alignment and code generation energy savings models. 
\item \exhaustTool for \salssa plus alignment models and synthesis per alignment model. We select the merged function version that saves the most energy.
\item Concat-style to evaluate resource sharing. We use this configuration as the baseline for all our experiments.  This is a more significant baseline than prior work in CGMA-generation~\cite{Brumar2021accelmerger,mokri2020early}, which only demonstrates improvements with respect to non-merged accelerators without considering their potential to use resource sharing across the boundries of the merged functions. For this reason, we dedicate subsection~\ref{sec:resourceSharingResults} to characterizing the resource sharing potential of our concatenation-based baseline configuration.
\end{itemize}

\subsection{HLS and Physical Design}

The merged and original (unmerged) functions are transformed into RTL modules using Bambu-HLS~\cite{bambu}. The RTL modules are further lowered (physical design) via technology mapping, place-and-route, and bitstream generation using Xilinx's Vivado 2023.1~\cite{Vivado} tool via the synthesis and implementation steps. We are using the device \texttt{xc7z045ffg900-2} for all the designs with the \texttt{synth\_design -directive AreaOptimized\_high} and \texttt{power\_opt\_design} Vivado directives and the \texttt{BAMBU-AREA} high-level synthesis flag to achieve fine-grained merging and classical resource sharing through state-of-the-art HLS and synthesis flows.

We choose Bambu-HLS since it provides a flexible LLVM interface that allows aggressive interprocedural optimizations such as CGFM before the HLS step. By contrast,  the Vitis HLS tool only allows finer-granularity optimization on the LLVM IR before HLS and fails to synthesize merged functions. Bambu-HLS has also been shown to be competitive in terms of PPA trade-offs, performing better on average than a commercial tool and comparably to other open source tools~\cite{bambuSurvey}. Bambu HLS is actively maintained, benefiting from the latest HLS and resource sharing research. We also show experimentally the baseline resource sharing benefits for Bambu-HLS in Section~\ref{sec:resourceSharingResults}. 

\section{Results}
\label{sec:results}

\begin{figure*}

 \centering
\includegraphics[width=0.9\linewidth]
	{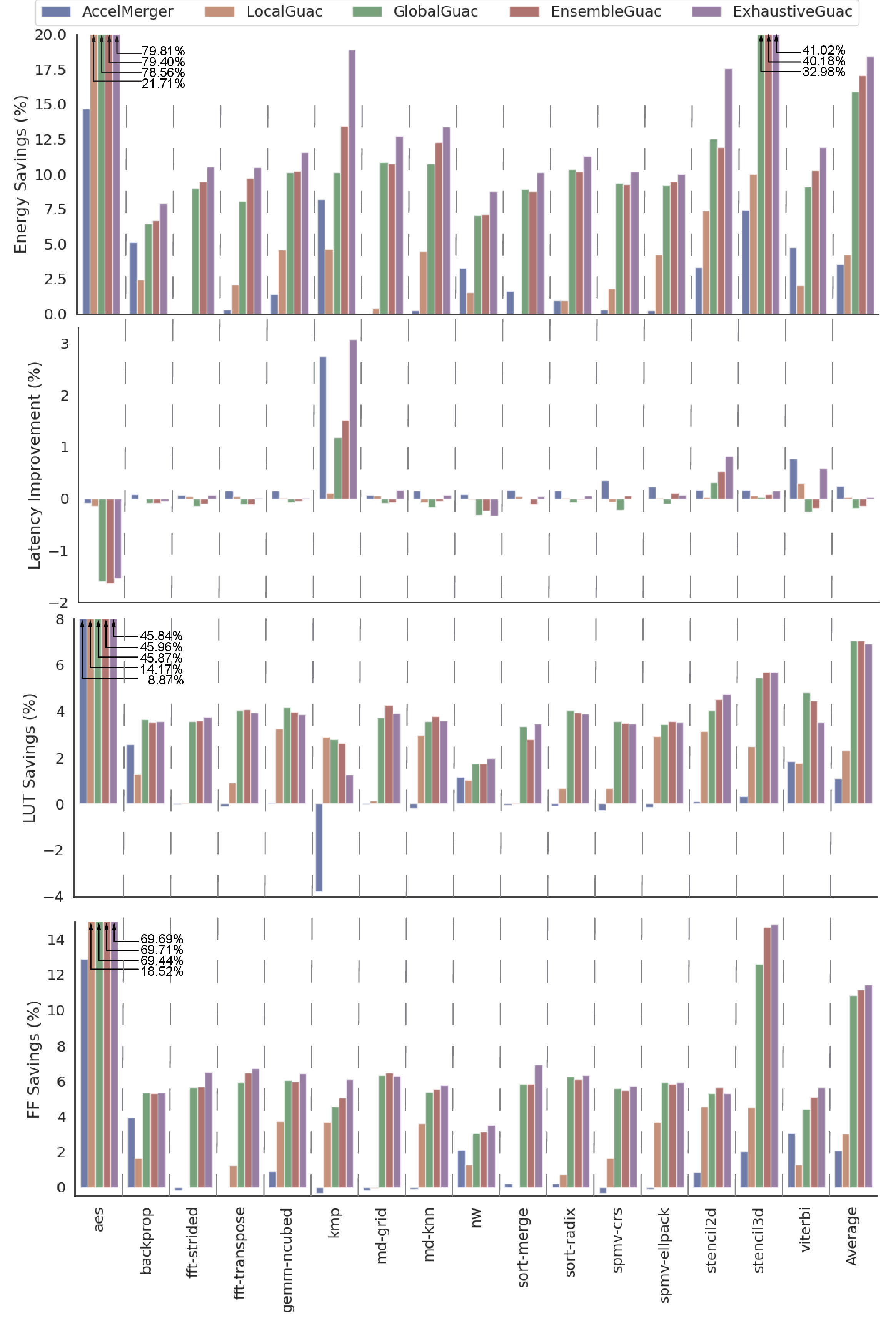}
 \caption{Energy, performance and area analysis with all the approaches. See Subsection~\ref{sec:configs} for a description of the merging configurations.}
 \label{fig:ppa}

\vspace*{-1.0em}
\end{figure*}

\begin{figure}
\centering
\begin{minipage}{.48\textwidth}
  \centering
  \includegraphics[width=0.95\linewidth]{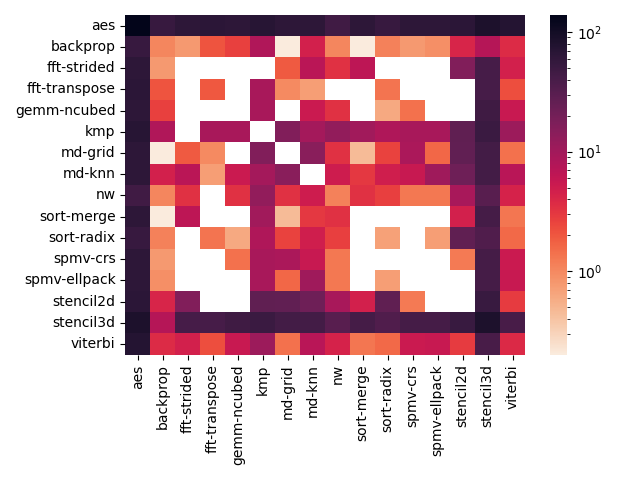}
  \captionof{figure}{The energy improvement (\%) of the \exhaustTool configuration with respect to concatenated accelerators benefiting from resource sharing. We use the heatmap to identify application similarity. Darker colors represent higher savings.}
  \label{fig:exhaustive_energy_savings_heat}
\end{minipage}
\hfill
\begin{minipage}{.48\textwidth}
  \centering
  \includegraphics[width=0.95\linewidth]{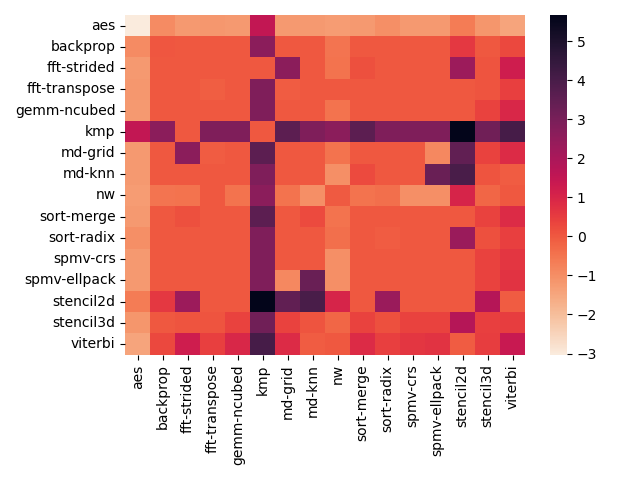}
  \captionof{figure}{Latency improvement (\%) of the \exhaustTool configuration with respect to concatenated accelerators benefiting from resource sharing. We use the same averaging of results for functions belonging to rows and columns as in Figure~\ref{fig:exhaustive_energy_savings_heat} but measuring $I_{CGMA, Latency}(f_1, f_2)$}
  \label{fig:exhaustive_latency_ovh_heat}
\end{minipage}
\vspace*{-1.0em}
\end{figure}

The results we present in this section represent both the intra- and inter-application merging benefits of \toolname. Each application contains functions that we merge against functions in other applications to evaluate the strength of the merging techniques on as many candidates as possible. We merge only the top-level function and its callees, filtering MachSuite-specific boilerplate functions. All merged functions are checked for correctness through co-simulation. We summarize the benefits corresponding to merges within the application and functions in other applications in the MachSuite benchmark suite.

\subsection{\toolname Area and Energy Savings}
\label{sec:areaEnergyBenefits}

We use Figure \ref{fig:ppa} to analyze the impact of the four \toolname configurations and \accelmerger in terms of area, energy, and latency.  This figure summarizes both intra- and inter-application merging results for each application. This means that we are showing the average improvement when merging functions from a fixed application against functions in all other applications. We will later provide the broken-down improvements per pair of applications. All the results in this subsection are normalized with respect to concatenated accelerators followed resource sharing. Later in Subsection~\ref{sec:resourceSharingResults} we will show a characterization of the baseline merged accelerators with resource sharing and concatenation.

Our evaluation indicates \globalTool achieves average improvements of 7.07\% in LUTs and 10.84\% in FFs, with an overhead 0.1\% in DSPs. Exploiting the similarities better between applications through SSA global alignments and the usage of more DSPs in \globalTool leads to a significant increase in power and energy savings. The average energy saving in Figure~\ref{fig:ppa} for \globalTool is 15.91\%. Despite the latency overhead for some applications, \globalTool yields important savings in energy consumption. 

As for model-based improvements, \modelTool further improves average energy savings to 17.08\%, and \exhaustTool yields an average of 18.4\% energy savings. If the \toolname user has the time budget to run HLS with the three cost models, \exhaustTool is better. If tool speed is important, \modelTool runs at a speed within 0.01\% of \globalTool. This is the effect of the selection model taking only 0.01\% of HLS and physical design time. Overall, \modelTool is our recommended CGMA technique. It can increase energy benefits with respect to \globalTool for aes, fft-transpose, kmp, md-knn, and viterbi. For all applications, however, \exhaustTool further improves energy savings, indicating future work opportunities for both function alignment and post-code-generation models. Applications like kmp and stencil2d highlight opportunities for more accurate post-code-generation models due to the gap in energy savings between \exhaustTool and \modelTool. Alignment algorithms that can benefit directly from function-level models rather than instruction-level models could allow all the \toolname configurations to derive even higher improvements.

Even though \exhaustTool and \modelTool bring great energy savings, as they optimize for energy savings through the post-code-generation model and directly through exhaustive synthesis with all the alignment models, they sometimes do so at a penalty in area. This can be observed for the viterbi application in LUT-savings being lower for \exhaustTool than \modelTool, which is lower than \globalTool. However in average \exhaustTool has negligible area overheads (less than 0.2\%) and \modelTool has no area overhead. In term of Flip Flops (FF plot), the \modelTool and \exhaustTool only improve the energy savings outcome. In terms of DSPs \exhaustTool has an average overhead of 0.13\% derived from non-zero DSPs overhead in Stencil2d (2.67\%), and negligible (less than 1\%) overheads in md-grid, nw, and sort-radix. \modelTool has an average DSPs overhead of 0.07\% with heaviest usage increase in stencil2d too (1.32\%) and negligible overhead in sort-radix. For all other combinations of application and CGFM technique, the DSP overheads are 0\% (since in most cases DSP overheads are 0\% we fully describe those area results in this paragraph instead of allocating a figure for these results).

Interestingly, for the \globalTool configuration, the top energy savings are achieved for aes, stencil3d and stencil2d at 67.57\%, 31.46\%, and 11.91\%. We see that backdrop, kmp, nw, merge-sort and viterbi benefit more from \accelmerger-style merges than \localTool-style merges, favoring global optimums across different basic blocks, rather than prioritizing branch merging and phi node handling (which is \localTool's main benefit). For more control-flow intense applications like stencil2d, gemm-ncubed and, stencil2d \localTool works better due to the branch-merging characteristics being more important than global merging. In general, we see that global merging on the SSA-representation is much more powerful than more recent CGMA-generating techniques relying on local merges (\localTool).

In Figure~\ref{fig:exhaustive_energy_savings_heat} we see the breakdown of energy savings inter- and intra-application. For each cell represented by $app_{row}$ and $app_{col}$ we are using the average improvement in energy savings produced by a pair of functions belonging to the two applications identifying the row and the column $\sum_{\forall f_1 \in app_{row}, f_2 \in app_{col}} \frac{I_{CGMA, Energy}\left(f_1, f_2\right)}{|app_{row}|*|app_{col}|}$. $app_{row}$ and $app_{col}$ are  sets of functions and we use $|app_{col\_or\_row}|$ to get the size of the set. We see that aes and stencil3d merge well consistently against other applications. When comparing these results to the concatenation experiments in Subsection~\ref{sec:resourceSharingResults}, we will see that CGMAs uncover new possible merges, such as for kmp and nw, since resource sharing only derives modest energy savings from these applications.

\begin{wrapfigure} {r}{0.5\textwidth}

 \centering
\includegraphics[width=0.9\linewidth]
	{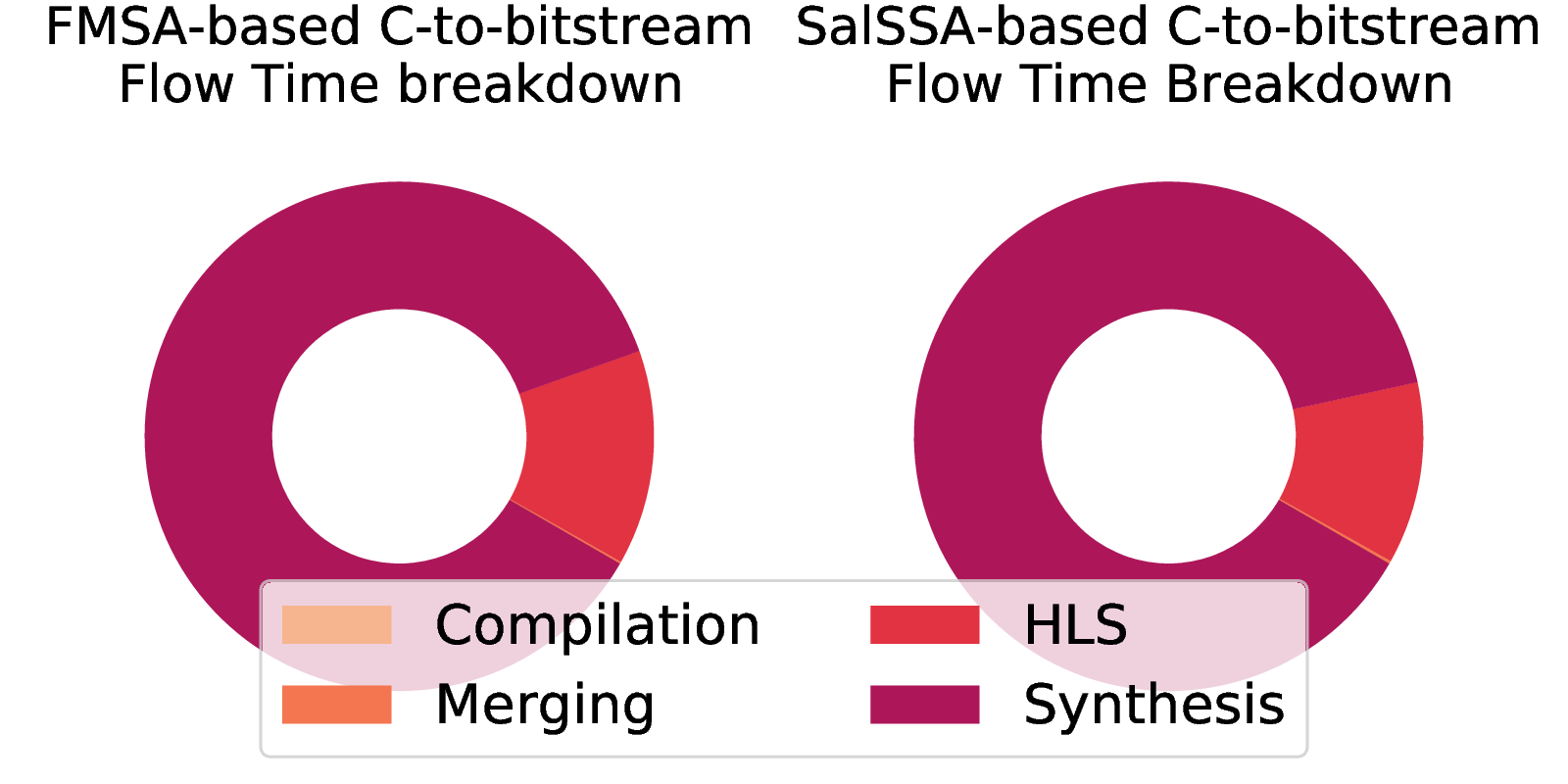}
 \caption{Total Synthesis Flow Time Breakdown for \accelmerger (left) and \modelTool (right). HLS and Synthesis dominate tool execution time. Compilation and Merging time is negligible.}
 \label{fig:timing_donut_plot}

\vspace*{-1.0em}
\end{wrapfigure}

 Figure~\ref{fig:exhaustive_latency_ovh_heat} shows latency improvement with respect to resource sharing when we break down the merges per pair of applications. We see that for most applications, latency improvement is 0 (orange)  or higher (represented with darker colors). Latency improves because CGFM, through the energy models used in this paper, targets energy improvements, and energy is proportional to latency. Also, area and power savings are traded off by HLS to reduce latency, which is reflected in a latency improvement of up to 6.1\% (represented in black). As we know from Figure~\ref{fig:ppa}, overall there are no latency overheads nor gains for the \exhaustTool.

We observe a negligible increase in overall HLS, synthesis and physical design time, with compilation from C/C++ to LLVM, standard optimizations (-O1) and merging  techniques comprising less than 0.3\% of total \modelTool flow time and less than 0.2\% of total \fmsa flow time, as shown in Figure~\ref{fig:timing_donut_plot}.  In this plot, we include running the ensemble model as part of the flow under the "Merging" concept. The model itself represents only 0.2\% of the overall flow.

Overall the code transformation time necessary for coarse-grained merging relative to the overall HLS and physical design time is low. This suggests that our approach is able derive significant energy savings at a very small execution time penalty with respect to \accelmerger and even with respect to approaches that are purely based on resource sharing.

\subsection{Resource Sharing baseline results}
\label{sec:resourceSharingResults}

\begin{table}[]
\begin{tabular}{ccccccc}
\textbf{App}                                                & \textbf{$I_{LUT}$} & \textbf{$I_{FF}$} & \textbf{$I_{DSP}$} & \textbf{$L_{ovh}$} & \textbf{$I_{Energy}$} \\\hline
\begin{tabular}[c]{@{}c@{}}aes\\ \end{tabular}         & 31.21        & 28.38       & 0.00         & 1.06                                                                                                                                                                      & 19.52                                                               \\\hline
\begin{tabular}[c]{@{}c@{}}back-\\ prop\end{tabular}       & 54.13        & 58.08       & 33.29        & 4.41                                                                                                                                                                      & 47.23                                                               \\\hline
\begin{tabular}[c]{@{}c@{}}fft\\ strided\end{tabular}       & 57.17        & 60.07       & 41.74        & 5.58                                                                                                                                                                      & 75.22                                                               \\\hline
\begin{tabular}[c]{@{}c@{}}fft\\ transpose\end{tabular}      & 33.14        & 31.70       & 17.53        & 1.90                                                                                                                                                                      & 33.20                                                               \\\hline
\begin{tabular}[c]{@{}c@{}}gemm\\ blocked\end{tabular}      & 60.99        & 65.50       & 41.74        & 10.88                                                                                                                                                                     & 58.28                                                               \\\hline
\begin{tabular}[c]{@{}c@{}}gemm\\ ncubed\end{tabular}         & 62.51        & 63.09       & 41.74        & 3.88                                                                                                                                                                      & 61.10                                                               \\\hline
\begin{tabular}[c]{@{}c@{}}kmp\\ \end{tabular}         & 11.81        & 12.00       & 0.00         & 4.54                                                                                                                                                                      & 9.98                                                                \\\hline
\begin{tabular}[c]{@{}c@{}}md\\ grid\end{tabular}         & 67.44        & 72.71       & 41.74        & 3.88                                                                                                                                                                      & 64.17                                                               \\\hline
\begin{tabular}[c]{@{}c@{}}md\\ knn\end{tabular}          & 59.12        & 65.48       & 41.74        & 3.88                                                                                                                                                                      & 53.75                                                               \\\hline
\begin{tabular}[c]{@{}c@{}}nw\\ \end{tabular}           & 9.34         & 8.00        & 0.00         & 14.51                                                                                                                                                                     & 14.20                                                               \\\hline
\begin{tabular}[c]{@{}c@{}}sort\\ merge\end{tabular}      & 19.88        & 26.74       & 0.00         & 18.99                                                                                                                                                                     & 14.55                                                               \\\hline
\begin{tabular}[c]{@{}c@{}}sort\\ radix\end{tabular}      & 24.80        & 33.07       & 0.00         & 6.99                                                                                                                                                                      & 39.04                                                               \\\hline
\begin{tabular}[c]{@{}c@{}}spmv\\ crs\end{tabular}        & 58.76        & 60.07       & 41.16        & 3.87                                                                                                                                                                      & 64.66                                                               \\\hline
\begin{tabular}[c]{@{}c@{}}spmv\\ ellpack\end{tabular}       & 58.90        & 60.34       & 41.16        & 3.87                                                                                                                                                                      & 48.84                                                               \\\hline
\begin{tabular}[c]{@{}c@{}}stencil\\ 2d\end{tabular}      & 187.43       & 249.51      & 71.64        & 8.62                                                                                                                                                                      & 126.05                                                              \\\hline
\begin{tabular}[c]{@{}c@{}}stencil\\ 3d\end{tabular}      & 5.47         & 8.18        & 0.00         & 9.00                                                                                                                                                                      & 10.17                                                               \\\hline
\begin{tabular}[c]{@{}c@{}}viterbi\\ \end{tabular} & 54.89        & 63.28       & 0.00         & 3.48                                                                                                                                                                      & 38.55                                                               \\\hline
Avg.                                                        & 50.41        & 56.83       & 24.32        & 6.43                                                                                                                                                                      & 45.79         
\end{tabular}
\caption{We use resource sharing and the Concat configuration as a strong baseline for CGFM-generated accelerators. In this table we show the savings that can be achieved through function concatenation followed by resource sharing with respect to applying resource sharing only locally in the original functions (no concatenation).}
\label{table:resourceSharing}
\vspace*{-2.2em}
\end{table}

\begin{figure}
\centering
\begin{minipage}{.48\textwidth}
  \centering
  \includegraphics[width=0.95\linewidth]{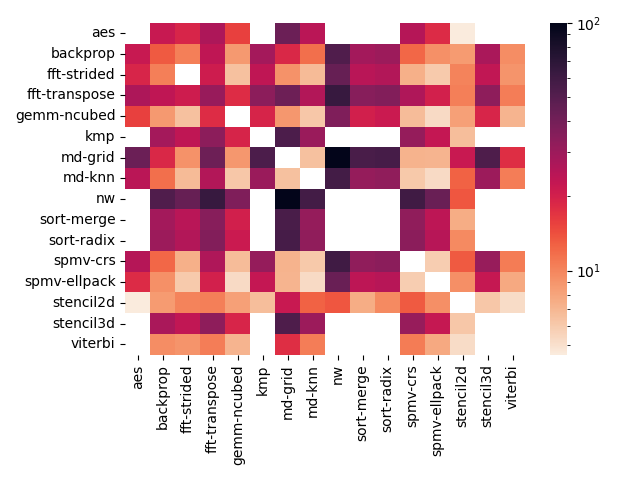}
  \captionof{figure}{The normalized (\%) energy savings $I_{Energy}$ for all pairs of merged applications when using function concatenation followed by HLS resource sharing. These results are computed with respect to the input accelerators. The input accelerators do not benefit from resource sharing across the two functions but they do benefit locally within the scope of their respective functions. The darker the color, the higher the energy savings.}
  \label{fig:concat_energy_savings_heat}
\end{minipage}
\hfill
\begin{minipage}{.48\textwidth}
  \centering
  \includegraphics[width=0.95\linewidth]{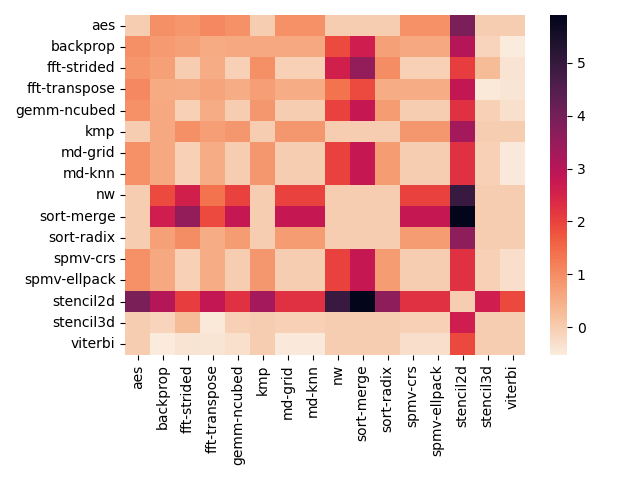}
  \captionof{figure}{Average percentage of latency overhead $L_{ovh}$ for all the pairs of merged applications for function concatenation followed by HLS resource sharing. Note that the HLS tool trades off area for latency in some cases resulting in latency improvements or losses. The maximum overhead is 5.9\% and the minimum is -0.52\% referring to latency improvements. Latency overheads are the results of adding multiplexers to handle the sharing of functional units.}
  \label{fig:concat_lat_improv_heat}
\end{minipage}
\vspace*{-2.0em}
\end{figure}
In Table~\ref{table:resourceSharing} we show the average resource, and energy savings percentages that can be achieved by merging accelerators with the Concat configuration. These results are relevant since the prior CGFM-based CGMA generation does not use such a strong baseline for their proposed techniques. We represent the average function merging savings among functions in the same application as well as across applications in this summary. For example the sort-merge row shows that this application saves an average of 19.88\% LUTs, 26.74\% FFs,  and 14.55\% in energy savings and it does not save any DSPs. In this case the power savings are larger than the energy savings due to an accelerator latency overhead of 18.99\%. As expected, resource sharing incurs a latency overhead by scheduling multiple operations to the same functional unit, leading to resource conflicts and also stretching the cycle time. Cycle time increases occur due to excessive routing towards the reused unit, but especially due to an increase in the FU input multiplexers that allow FUs to be shared. Scheduling the input multiplexer either in the cycle of the FU or in the preceding cycle can increase the cycle time, whereas scheduling the multiplexer in its own cycle leads to an overhead in register usage and cycle count.

Despite these trade-offs in FU and register reuse, concatenating functions and exposing them to resource sharing yields remarkable savings of 50.41\% in LUTs, 56.83\% in FFs (registers), 24.32\% DSPs, and 45.79\% in energy. This comes at a latency overhead of 6.43\% which is completely due to the cycle time increase, meaning that HLS does not find it profitable to schedule the multiplexers in a separate cycle. In some applications, trading speed for area results in highly non-linear savings in area and energy, e.g., in the case of stencil2d.

Figure~\ref{fig:concat_energy_savings_heat} provides a detailed view of the merged accelerators intra- (diagonal) and inter-applications. Since some applications contain a single function, the diagonal represents no savings since we do not allow identical function merging. In this work we are interested in merging similar accelerators. Whereas Table~\ref{table:resourceSharing} provides the summary of each row in this heat map, Figure~\ref{fig:concat_energy_savings_heat} breaks down those savings into the merges that caused the overall energy improvements.  

Most applications are similar enough to another application that inter-application merging generates energy savings. In particular, applications with rich control flow and a large variety of operation types (such as md) benefit the most from merging. Applications like nw and sort benefit from merging with a more constrained set of applications, whereas backprop, fft and gemm merge well across a wider variety of applications. Only backprop and fft-transpose benefit from intra-application merging. These larger applications contain functions large and similar enough to benefit from merging.

Figure~\ref{fig:concat_lat_improv_heat} shows that resource sharing can have a latency overhead of up to 5.9\%. In some cases, due to HLS trade-offs in area and latency, we observe negligible resource sharing latency improvements of up to 0.52\%.






\section{Related Work}
Most coarse grained merging for accelerator design is occurring in the context of manual code analysis and acceleration~\cite{suleiman2019navion, chen2014diannao, liu2015pudiannao, tambeLatest}. In this work, we improve automated coarse-grained merging to further motivate the use of coarse-grained accelerator merging.

The main related works in coarse-grained accelerator merging are AccelMerger~\cite{Brumar2021accelmerger} and ReconfAST\cite{mokri2020early}. They rely on analytical models to provide merging information early in accelerator design, while in this work we focus on a highly accurate C/LLVM-to-bitstream approach. ReconfAST implements the merging optimization in the C-front-end layer of clang (pre-LLVM), which makes it specific to a single high-level language, and less likely to benefit from mainstream compiler optimization research. Also, IRs are simpler and more amenable to optimizations like CGFM than high-level languages. For example, control flow in the IR is handled with branch and phi instructions, whereas at a high-level there are many control-related constructs, such as while, for, do while, if and switch statements. 

There is a large body of work on resource sharing that have shaped the area and power optimizations implemented in HLS and physical design tools~\cite{brisk2004area, venkatesh2011qscores, stitch, venkatesh2010conservation, lam2009rapid, moreano2005efficient, cong2008pattern}. 

Even though works discusses PPA estimation~\cite{modelOlder, modelfccm, zuo2017accurate, Brumar2021accelmerger, poem, hung2016kapow}  these models are rarely used for compiler optimization and not with the goal of interprocedural optimization aimed at CGMA generation. Cycle-accurate accelerator simulation literature has used linear operation-level optimizations but coarse grained accelerator merging is out of the scope of these works~\cite{gem5salam,shao2014aladdin}. CGFM-based CGMA accelerator works~\cite{mokri2020early, Brumar2021accelmerger} by contrast contain no contributions on models either at alignment time nor at post-code-generation time.


\fmsa, \hyfm and \salssa~\cite{rocha2019function, rocha21hyfm, rocha2020effective} are LLVM compilation transformations aimed code reduction. These techniques use the CPU ISA (linear) model for code size 
out of the box to guide the Needleman-Wunsch algorithm in finding an optimal alignment. Since LLVM lacks FPGA cost models, we implement our own cost models and replace candidate selection. Since HLS stages are much more complicated than mapping IR to some back-end CPU ISA, we use an ensemble of models emphasizing either control, data or memory instructions for merging.

\section{Conclusion}
This paper focuses on optimizing accelerator merging techniques and models for higher area and energy savings.
Our evaluation indicates average improvements in LUTs of 7.04\%, 11.17\% in FFs, with a penalty of 0.07\% in DSPs and energy savings of 17.08\%, respectively, with respect to a strong resource-sharing baseline. This represents a significant improvement with respect to state-of-the-art CGFM-based  CGMA generation, which improves energy savings by only 7.38\%, LUTs by 1.06\%, and FFs by 2.01\%. Finally, our post-code-generation and alignment models bring our solution \modelTool close to an exhaustive configuration \exhaustTool, which achieves 18.4\% energy savings and similar area improvements. These results motivate future work on enabling more transition of compiler techniques to the research field of HLS and work on enabling function level instead of operation level models in the alignment algorithms.

\bibliographystyle{ACM-Reference-Format}
\bibliography{sample-base}










\end{document}